\documentclass[10pt]{IEEEtran}%peerreview, eps version
\IEEEoverridecommandlockouts
\usepackage{algpseudocode}
\usepackage{algorithm}
\usepackage{bm}
\usepackage{amsfonts}
\usepackage{amssymb}
\usepackage{amscd}
\usepackage{graphics}
\usepackage[demo]{graphicx}
\usepackage{epsfig}
\usepackage{caption}
\usepackage{subcaption}
\usepackage[cmex10]{amsmath}
\usepackage{array}
\usepackage{eqparbox}
\usepackage{flushend}
\usepackage{fancyhdr}
\usepackage{titling}
\usepackage{enumerate}
\usepackage{float}
\usepackage{hyperref}
\usepackage{stackrel}

\DeclareMathAlphabet{\mathbsf}{OT1}{cmss}{bx}{n}% bold sans serif
\DeclareMathAlphabet{\mathssf}{OT1}{cmss}{m}{sl}% slanted sans serif
\DeclareMathAlphabet{\mathcsf}{OT1}{cmss}{sbc}{n}% condensed sans serif

\newcommand{\ie}{{\em i.e.}}

\newcommand{\iid}{i.i.d.}

\newcommand{\secref}[1]{Section~\ref{#1}}
\newcommand{\figref}[1]{Fig.~\ref{#1}}
\newcommand{\tabref}[1]{Table~\ref{#1}}

\makeatletter
\def\blfootnote{\xdef\@thefnmark{}\@footnotetext}
\makeatother

\newtheorem{theorem}{Theorem}[section]

\newtheorem{note}[theorem]{Note}

\newcommand{\qed}{\nobreak \ifvmode \relax \else
      \ifdim\lastskip<1.5em \hskip-\lastskip
      \hskip1.5em plus0em minus0.5em \fi \nobreak
      \vrule height0.75em width0.5em depth0.25em\fi}

\def\BibTeX{{\rm B\kern-.05em{\sc i\kern-.025em b}\kern-.08em
    t\kern-.1667em\lower.7ex\hbox{E}\kern-.125emX}}

\usepackage{newlfont}

\begin{document}
\title{An Electrical Structure-Based Approach to PMU Placement in the Electric Power Grid}
\author{\IEEEauthorblockN{K. G. Nagananda, Shalinee Kishore, \emph{Member, IEEE} and Rick S. Blum, \emph{Fellow, IEEE}}\thanks{K. G. Nagananda is with People's Education Society University, Bangalore, INDIA, E-mail: \texttt{kgnagananda@pes.edu}. Shalinee Kishore and Rick S. Blum are with Lehigh University, Bethlehem, PA, USA. E-mail: \texttt{\{skihsore, rblum\}@lehigh.edu}}}

\pagenumbering{gobble}
\date{}
\setlength{\droptitle}{-0.55in}
\maketitle

\begin{abstract}
The phasor measurement unit (PMU) placement problem is revisited by taking into account a stronger characterization of the electrical connectedness between various buses in the grid. To facilitate this study, the placement problem is approached from the perspective of the \emph{electrical structure} which, unlike previous work on PMU placement, accounts for the sensitivity between power injections and nodal phase angle differences between various buses in the power network. The problem is formulated as a binary integer program with the objective to minimize the number of PMUs for complete network observability in the absence of zero injection measurements. The implication of the proposed approach on static state estimation and fault detection algorithms incorporating PMU measurements is analyzed. Results show a significant improvement in the performance of estimation and detection schemes by employing the electrical structure-based PMU placement compared to its topological counterpart. In light of recent advances in the electrical structure of the grid, our study provides a more realistic perspective of PMU placement in the electric power grid.
\end{abstract}

\vspace{-0.15in}
\section{Introduction}\label{sec:introduction}
In the classic setting, the problem of placing PMUs within an electric power system is divided into two parts: 1) obtaining the optimal or minimum number of PMUs; and 2) finding the optimal locations to install these PMUs on the grid.  In both stages, the optimizations are typically done with the goal to have either complete or incomplete network observability either in the presence/absence of  zero injection measurements \cite{Nuqui2005}. Zero injection measurements are present when the power system has nodes without generation or load. The placement problem is formulated as an integer linear program, comprising the adjacency (or connectivity) matrix of the grid, to obtain the minimal set of PMUs and their optimal locations.

Various strategies have been investigated to address the PMU placement problem. For instance, in \cite{Baldwin1993}, the placement was provided by a spanning measurement subgraph, and the minimal PMU set was obtained using a dual search algorithm incorporating modified bisecting simulated-annealing-based searches for complete network observability. In \cite{Milosevic2003}, a Pareto-optimal solution was obtained using non-dominated sorting genetic algorithm, which turned out to be nonlinear in the presence of power injection measurements. In \cite{Xu2004}, the problem was formulated as an integer program to include conventional power flow and injection measurements in addition to PMU measurements for maximum network observability. In \cite{Chen2006}, the objective was to achieve bad data detection during state estimation using optimal PMU placement. In \cite{Gou2008}, \cite{Gou2008a}, optimal placements were obtained by formulating the problem as a simple integer linear program to achieve standard objectives.

In \cite{Zhang2010}, the optimal placement was linked with power system dynamic state estimation. A method for evaluating a specific placement configuration, when there were multiple solutions, was also proposed. In \cite{Azizi2012}, the problem was solved using an exhaustive search-based method with complete network observability, and the state estimation implemented on such a placement was shown to be linear. A convex relaxation method was employed in \cite{Kekatos2012} to optimize the placement based on estimation-theoretic criteria under the Bayesian framework. In \cite{Li2013}, an information-theoretic measure, namely, mutual information (MI) was employed to address the PMU placement problem, where the objective was to maximize the MI between PMU measurements and power system states to obtain highly ``informative'' PMU configurations.

Studies in the aforementioned work were based on the topology or node connectivity (degree distribution) of the grid. However, advances in complex networks have questioned the veracity of characterization of the power grid using its topological structure \cite{Dorfler2010}. Subsequently, several demerits of the topology-based analysis were uncovered. For instance:
\begin{enumerate}[(1)]
\item Studies showed that, for many real-world power grids, characterizing the network structure using topology alone was suboptimal, and had implications on node synchronization and performance of the network \cite{Wu1995} - \nocite{Wu2005}\cite{Atay2006}.
\item It was reported that grids in different geographical regions had different degree distributions, leading to varied topological structures. Furthermore, it was also shown that different model-based analyses of the same grid had resulted in different topological structures \cite{Albert2004}, \cite{Chassin2005}.
\item It was shown that the topological structure did not comprehensively capture the underlying physical laws (Ohm's and Kirchoff's) that govern the electrical connections or flows between network components \cite{Cotilla-Sanchez2012}.
\item In the context of PMU placement, the topology-based approach provides a weak characterization of the electrical influence between various components in the grid, since it does not take into account the effect of bus voltage magnitudes and phasor angles. Also, owing to the the time-varying (albeit slowly) nature of the networks' connectivity, an optimal PMU placement (minimum number and locations) apposite to a particular instant in time might not hold for a different time instant.
\end{enumerate}
To overcome the aforementioned drawbacks of the topological structure of the grid, authors of \cite{Cotilla-Sanchez2012} studied the \emph{electrical structure}, which accounted for the sensitivity between power injections and nodal phase angle differences between buses in the grid, to provide a more comprehensive characterization of the connectedness between various network components.

Motivated by these observations, the PMU placement problem is revisited in this paper from the perspective of the electrical structure of the grid. To the best of the our knowledge, this is the first instance of the placement problem being addressed from the perspective of the electrical structure, which provides a much stronger characterization of the electrical influence between various buses in the gird. As the results demonstrate, a larger number of PMUs are required when compared with the number required when using the topological structure to meet the objective of complete network observability in the absence of zero injection measurements. However, in light of the aforementioned arguments which favor the electrical structure of the grid, our results provide a more realistic perspective of PMU placement. The studies in this paper are topical especially with the proposition by the electric utilities to install large numbers of PMUs on the grid.

The implication of the proposed approach on the performance of (i) static state estimation and (ii) fault detection algorithm to detect changes in the susceptance parameters of the grid is analyzed. The detection problem is formulated using a linear errors-in-variables (EIV) model, and a generalized likelihood ratio test (GLRT) based on the total maximum likelihood (TML) methodology is presented. Results show a significant improvement in the performance of both estimation and detection schemes by employing the electrical structure-based PMU placement compared to its topological counterpart.

In \secref{sec:review_elecstruc}, a review of the electrical structure of the grid is provided to build the necessary framework to be used in the rest of the paper. In \secref{sec:main_results}, the placement problem is formulated and the results are tabulated for several IEEE bus systems. State estimation and fault detection frameworks with PMU measurements are presented in \secref{sec:static_estimation_fault detection}. \secref{sec:simulation} comprises experimental results and related discussion. We conclude the paper in \secref{sec:conclusion}.
\vspace{-0.075in}
\section{Electrical Structure of the Power Network}\label{sec:review_elecstruc}
In this section, the electrical structure of the grid from the perspective of complex networks is reviewed. The concept of resistance distance is introduced, and the procedure to derive the binary connectivity matrix of the power grid is presented (see \cite[Section III]{Cotilla-Sanchez2012} for further details).

The sensitivity between power injections and nodal phase angles differences can be utilized to characterize the electrical influence between network components. The electrical structure of the power network can then be understood by measuring the amount of electrical influence that one component has on another in the network. The measurement of this electrical influence necessitates a metric system. Mathematically, this can be accomplished by first deriving the sensitivity matrix, which can be obtained by well-established methods. The complement of the sensitivity matrix is called the distance matrix, whose entries quantify the electrical influence that each component has on the other - zero value indicates that two components are perfectly connected, while a large number indicates that the corresponding components have negligible electrical influence on each other. This electrical distance was proved to be a formal distance metric (see \cite[Appendix]{Cotilla-Sanchez2012}), and was employed to address various problems in the grid.

Another method to measure the electrical influence between network components is to derive the resistance distance \cite{Klein1993}, which is the effective resistance between points in a network of resistors. Consider a network with $N$ nodes, described by the conductance matrix $\bm{G}$. Let $V_j$ and $g_{ij}$ denote the voltage magnitude at node $j$ and the conductance between nodes $i$ and $j$, respectively. The current injection at node $i$ is given by $I_i = \sum_{j=1}^{N}g_{ij}V_j$. $\bm{G}$ acts as a Laplacian matrix to the network, provided there are no connections to the ground, {\ie}, if $\bm{G}$ has rank $N-1$. The singularity of $\bm{G}$ can be overcome by letting a reference node $r$ have $V_r = 0$. The conductance matrix associated with the remaining $N-1$ nodes is full-rank, and thus $\bm{V}_k = \bm{G}^{-1}_{kk}\bm{I}_k, k \neq r$, where $\bm{G}^{-1}_{kk}$ is the sub-matrix of $\bm{G}$ associated with the non-reference nodes, and $\bm{V}_k$ and $\bm{I}_k$ denote the vectors of voltage magnitudes and current injections, respectively, at the non-reference nodes $k \neq r$.

Let the diagonal elements of $\bm{G}^{-1}_{kk}$ be denoted $g^{-1}_{kk}$, $\forall k$, indicating the change in voltage due to current injection at node $k$ which is grounded at node $r$. The voltage difference between a pair of nodes $(i,j)$, $i\neq j\neq r$, is $e(i,j) = g^{-1}_{ii} + g^{-1}_{jj} - g^{-1}_{ij} - g^{-1}_{ji}$, indicating the change in voltage due to injection of $1$ Ampere of current at node $i$ which is withdrawn at node $j$. $e(i,j)$ is called the resistance distance between nodes $i$ and $j$, and describes the sensitivity between current injections and voltage differences. Letting $\boldsymbol{\Gamma} \triangleq \text{diag}(\bm{G}^{-1}_{kk})$ and $\boldsymbol{1}$ denote the vector of all ones, we have
\begin{eqnarray}
\bm{E}_{kk} &=& \boldsymbol{1}\boldsymbol{\Gamma}^{\mathrm{T}} + \boldsymbol{\Gamma}\boldsymbol{1}^{\mathrm{T}} - \bm{G}^{-1}_{kk} - \left[\bm{G}^{-1}_{kk}\right]^{\mathrm{T}},\label{eq:matrix_form1}\\
\bm{E}_{rk} &=& \boldsymbol{\Gamma}^{\mathrm{T}},\label{eq:matrix_form2}\\
\bm{E}_{kr} &=& \boldsymbol{\Gamma}.\label{eq:matrix_form3}
\end{eqnarray}
\eqref{eq:matrix_form1} - \eqref{eq:matrix_form3} holds for $\forall k \neq r$. The resistance distance matrix $\bm{E}$, thus defined, possesses the properties of a metric space \cite{Klein1993}.

To derive the sensitivities between power injections and phase angles, we start with the upper triangular part of the Jacobian matrix obtained from the power flow analysis, for the distance matrix to be real-valued:
\begin{eqnarray}\label{eq:upper_jacobian}
\Delta \bm{P} = \left[\frac{\partial P}{\partial \theta}\right]\Delta \theta + \left[\frac{\partial P}{\partial |V|}\right]\Delta |V|.
\end{eqnarray}
The matrix $\left[\frac{\partial P}{\partial \theta}\right]$ will be used to form the distance matrix by assuming the voltages at the nodes to be held constant, {\ie}, $\Delta |V| = 0$. It was observed that $\left[\frac{\partial P}{\partial \theta}\right]$ possesses most of the properties of a Laplacian matrix (see \cite{Cotilla-Sanchez2012}). By letting $\bm{G} = \left[\frac{\partial P}{\partial \theta}\right]$, the resulting distance matrix $\bm{E}$ measures the incremental change in phase angle difference between two nodes $i$ and $j$, $(\theta_i - \theta_j)$, given an incremental average power transaction between those nodes, assuming the voltage magnitudes are held constant. It was proved in \cite[Appendix]{Cotilla-Sanchez2012} that $\bm{E}$, thus defined, satisfies the properties of a distance matrix, as long as all series branch reactances are nonnegative.

For a power grid with $N$ buses, the distance matrix $\bm{E}$ translates into an undirected graph with $N(N-1)$ weighted branches. In order to describe a grid with an undirected network without weights, one has to retain the $N$ buses, but replace the $K$ branches with $K$ smallest entries in the upper or lower triangular part of $\bm{E}$. This results in a graph of size $\{N,M\}$ with edges representing electrical connectivity rather than direct physical connections. The connectivity matrix $\bm{C}$ of this graph is obtained by setting a threshold, $\tau$, adjusted to produce exactly $K$ branches in the network:
\begin{eqnarray}
\bm{C}:
\begin{cases}
c_{ij} = 1, ~\forall e(i,j) < \tau,\\
c_{ij} = 0, ~\forall e(i,j) \geq \tau
\end{cases}
\label{eq:adjcency_matrix}
\end{eqnarray}
To obtain the threshold $\tau$ in \eqref{eq:adjcency_matrix}, we first consider the upper triangular part of the matrix $\bm{E}$, and sort the elements in descending (or ascending) order. We then pick a number of elements equal to the number of branches in the given power network. For instance, in the IEEE-14 bus network having 20 branches, we pick the top 20 sorted elements from the upper triangular part of the matrix $\bm{E}$. In this paper, the binary connectivity matrix $\bm{C}$ is derived for several standard IEEE test bus systems for use in the PMU placement problem.
\vspace{-0.1in}
\section{Problem formulation and main results}\label{sec:main_results}
The PMU placement problem is formulated and addressed from two different perspectives: one based on the topological structure of the grid and the other based on its electrical structure. The problem is setup as an integer linear program to achieve complete network observability in the absence of zero injection measurements. The extension to the case of incomplete network observability (see \cite{Nuqui2005}) is straightforward and, therefore, is not considered in this paper.

Consider a power network with $N$ buses and $K$ branches. Let $i$ and $j$ to denote the bus indices, $i, j\in \{1,\dots,N\}$, and let $\bm{C}$ denote the $N\times N$ binary connectivity matrix. Let $\bm{d}$ be an $N\times 1$ binary decision variable vector defined by
\begin{eqnarray}\label{eq:binarydecision_vector}
d_i = \begin{cases}
1, ~\text{if a PMU is installed at bus}~i,\\
0, ~\text{otherwise},
\end{cases}
\end{eqnarray}
where $i=1,\dots,N$, and $\bm{b}$ a unit vector of dimensions $N\times 1$. The PMU placement problem for complete network observability and without conventional measurements is formulated as follows \cite{Gou2008}, \cite{Gou2008a}:
\begin{eqnarray}\label{eq:pmuplacement}
\nonumber \min \sum_{i=1}^{N}d_i\\
\text{such that}~ \bm{C}\bm{d} \succeq \bm{b}.
%\nonumber d_i \in \{0,1\}
\end{eqnarray}

For PMU placement based on the topological structure of the power network, we consider the existing approach of deriving the binary connectivity matrix $\bm{C}$ directly from the bus admittance matrix. The entries of the bus admittance matrix are transformed into binary form, and used in the problem setup \eqref{eq:pmuplacement}. The entries of $\bm{C}$ are given by
\begin{eqnarray}\label{eq:topological_A}
\bm{C}:
\begin{cases}
c_{ij} = 1, ~\text{if}~i=j,\\
c_{ij} = 1, ~\text{if}~i~\text{and}~j~\text{are connected},\\
c_{ij} = 0, ~\text{if}~i~\text{and}~j~\text{are not connected}.
\end{cases}
\end{eqnarray}
The entries $c_{ij}$ characterize the electrical influence between the buses $i$ and $j$ based on the degree distribution of the grid. For the electrical structure-based approach, the connectivity matrix $\bm{C}$ is given by \eqref{eq:adjcency_matrix}.

The bus and branch data required to derive the bus admittance and power-flow Jacobian matrices were obtained using MATPOWER \cite{Zimmerman2011}. For the topology-based approach, the connectivity matrix is computed using \eqref{eq:topological_A}, which is employed in \eqref{eq:pmuplacement} to obtain the placement soution. Similarly, for the electrical structure-based approach, the connectivity matrix is obtained using \eqref{eq:adjcency_matrix} as described in \secref{sec:review_elecstruc}. For both approaches, the criterion for the optimization problem was to achieve complete network observability, without conventional measurements. The binary integer programming tool of Matlab was used to solve the problem defined by \eqref{eq:pmuplacement}. %
\begin{table}[h]
\centering
\begin{tabular}{ | c | c | c |}\hline
    IEEE bus system & Topological structure & Electrical structure \\ \hline
    9 & 3 &  4\\ \hline
    14 & 4 & 7 \\ \hline
    30 & 10 & 20 \\ \hline
    39 & 13 & 24 \\ \hline
    57 & 17 & 39 \\ \hline
    118 & 32 & 98 \\ \hline
    162 & 43 & 131 \\ \hline
    \end{tabular}
    \caption{Minimum number of PMUs based on topological and electrical structures for IEEE test bus systems}
    \label{tab:bus_pmu}
\end{table}

The main result of the paper is tabulated in \tabref{tab:bus_pmu}. It can be seen that for each test bus system, the minimum number of PMUs obtained using the electrical structure is higher than that obtained using the topological structure, to meet the objective of complete network observability in the absence of zero injection measurements. Though this result is unfavorable from an economic standpoint, the arguments presented in \secref{sec:introduction} clearly support the employment of more PMUs for optimal system operation. Furthermore, the optimal locations of the PMUs are also different for these two approaches. For instance, for the 14 bus system, the topology-based approach suggests to place the 4 PMUs on buses numbered 2, 6, 7 and 9, which is in agreement with \cite{Gou2008}, \cite{Gou2008a}. Whereas, the electrical structure-based placement scheme suggests to place the 7 PMUs on buses numbered 1, 3, 8, 11, 12, 13 and 14. A similar behavior was noticed for the other bus networks, however, in the interest of space the observations are not reported here. Next, the advantages offered by the electrical structure-based PMU placement over the topology-based approaches is demonstrated via state estimation and fault detection algorithms.

\vspace{-0.1in}
\section{Applications using PMU measurements}\label{sec:static_estimation_fault detection}
\subsection{Static state estimation}\label{subsec:state_estimation}
In this section, results of incorporating PMU measurements into a static state estimation algorithm is presented. The minimal number of PMUs and their optimal locations are decided by the solution to the PMU placement problem \eqref{eq:pmuplacement}. The implications of both topology-based and electrical structure-based approaches to PMU placement on the performance of state estimation are discussed. In static state estimation, the control center accumulates a single snapshot of measurements, collected from meter readings and other available information from supervisory control and data acquisition systems (SCADA), to provide estimates of the system state. The measurement model is first presented, into which the PMU measurements are augmented.

For a power grid comprising $N$ buses, the system state vector to be estimated is $\bm{s} = [\theta_2,\dots,\theta_{N}, V_1,\dots,V_{N}]$. The control center accumulates a measurement vector $\bm{y} \in \mathbb{R}^{L \times 1}, L > 2N-1$ to estimate the state vector $\bm{s}$. The measurements consist of active and reactive power flows and power injections at the buses. The active and reactive power flows between buses $i$ and $j$ are denoted $P_{ij}$ and $Q_{ij}$, respectively, while $P_{i}$ and $Q_{i}$ denote the active and reactive power injections at bus $i$, respectively. The following relations characterize the measurement function \cite{Kotiuga1982}:
\begin{eqnarray}
\nonumber P_{ij} &=& V_{i}V_{j}\gamma_{ij}\cos (\theta_{i}-\theta_{j}+\alpha_{ij}) \\ && - V_{i}^2\gamma_{ij}\cos (\alpha_{ij}) + V_{i}^2\gamma_{sj}\cos (\alpha_{sj}),\label{eq:measure_function1}\\
\nonumber Q_{ij} &=& V_{i}V_{j}\gamma_{ij}\sin (\theta_{i}-\theta_{j}+\alpha_{ij})\\ && - V_{i}^2\gamma_{ij}\sin (\alpha_{ij}) + V_{i}^2\gamma_{\mathrm{s}j}\sin (\alpha_{\mathrm{s}j}),\label{eq:measure_function2}\\
P_{i} &=& \sum_{j=1}^{B}V_{i}V_{j}\gamma_{ij}\cos (\theta_{i}-\theta_{j}+\alpha_{ij}),\label{eq:measure_function3}\\
Q_{i} &=& \sum_{j=1}^{B}V_{i}V_{j}\gamma_{ij}\sin (\theta_{i}-\theta_{j}+\alpha_{ij}),\label{eq:measure_function4}
\end{eqnarray}
where $\gamma_{ij}$ and $\alpha_{ij}$ are the magnitude and phase angle of the admittance from bus $i$ to bus $j$, respectively, and $\gamma_{sj}$ and $\alpha_{sj}$ are the magnitude and phase angle of the shunt admittance at bus $i$, respectively. $\gamma_{ij} = 0$ if the buses $i$ and $j$ are not connected, which models the topology constraints of the network. The vlaues $\gamma_{ij}$, $\alpha_{ij}$, $\gamma_{sj}$ and $\alpha_{sj}$ are assumed known.

The measurement model is given by $\bm{y} = \bm{f}(\bm{s}) + \bm{v}$, where $\bm{f}$ denotes the set of $L$ nonlinear functions specified by \eqref{eq:measure_function1} - \eqref{eq:measure_function4} relating the state and measurement variables, and $\boldsymbol{v}$ denotes the $L \times 1$ measurement noise vector assumed to be Gaussian with covariance matrix $\bm{\Upsilon}$. The Jacobian matrix $\bm{J}$ of the network is given by the partial derivative $\bm{J} \triangleq \frac{\partial(\bm{f}(\bm{s}))}{\partial(\bm{s})}$. The weighted least-squares (WLS) estimate of the state vector $\bm{s}$ at the $(t+1)^{\text{th}}$ iteration is $\hat{\bm{s}}_{t+1} = \hat{\bm{s}}_{t} + \bm{L}_{t}\bm{J}_t^{\mathrm{T}}\bm{\Upsilon}^{-1}\left(\bm{y} - \bm{f}(\bm{s}_{t})\right)$ (see\cite{AliAbur2004}), where the gain matrix is $\bm{L}_{t} \triangleq \left(\bm{J}_t^{\mathrm{T}}\bm{\Upsilon}^{-1}\bm{J}_t\right)^{-1}$. When $\bm{J}_t^{\mathrm{T}}\bm{\Upsilon}^{-1}\left(\bm{y} - \bm{f}(\bm{s}_{t})\right)$ attains a predefined value, the estimator is said to have converged, and the error covariance matrix of the estimate $\hat{\bm{s}}$ is given by $\bm{R}_{\hat{\bm{s}}} = \left(\bm{J}^{\mathrm{T}}\bm{\Upsilon}^{-1}\bm{J}\right)^{-1}$.

In \cite{Zhou2006}, a procedure was devised to incorporate the PMU measurements into the state estimation framework described above. By using the fact that the PMU is designed to record not only the voltage magnitude and phase angles at the bus where it is installed, but also current phasors of all lines incident on that bus, the measurements at the PMUs is defined by $\tilde{\bm{y}} = \left[V_{\text{real}}~V_{\text{img}}~I_{\text{real}}~I_{\text{img}} \right]$, where $V_{\text{real}}$, $V_{\text{img}}$, $I_{\text{real}}$ and $I_{\text{img}}$ are the real and imaginary parts of the voltage and current phasors, respectively, given by
\begin{eqnarray}
V_{i,\text{real}} &=& \left|V_{i}\right|\cos \theta_i, \label{eq:pmu_measure1}\\
V_{i,\text{img}} &=& \left|V_{i}\right|\sin \theta_i, \label{eq:pmu_measure2}\\
\nonumber I_{ij,\text{real}} &=& \left(\left|V_{i}\right|\cos \theta_i - \left|V_{j}\right|\cos \theta_j\right)g_{ij} - b_{i0}\left(\left|V_{i}\right|\sin \theta_i\right) \\ && - \left(\left|V_{i}\right|\sin \theta_i - \left|V_{j}\right|\sin \theta_j\right)b_{ij}, \label{eq:pmu_measure3}\\
\nonumber I_{ij,\text{img}} &=& \left(\left|V_{i}\right|\cos \theta_i - \left|V_{j}\right|\cos \theta_j\right)b_{ij}  + b_{j0}\left(\left|V_{i}\right|\sin \theta_i\right)\\ && + \left(\left|V_{i}\right|\sin \theta_i - \left|V_{j}\right|\sin \theta_j\right)g_{ij},\label{eq:pmu_measure4}
\end{eqnarray}
where $g_{ij}$ and $b_{ij}$ are the conductance and susceptance between buses $i$ and $j$, respectively. The measurement model for the PMU measurements is, therefore, given by $\hat{\bm{y}} = \tilde{\bm{f}}(\bm{s}) + \tilde{\bm{v}}$, where $\tilde{\bm{f}}$ is defined by \eqref{eq:pmu_measure1} - \eqref{eq:pmu_measure4}, and $\tilde{\bm{v}}$ is the PMU measurements noise vector assumed to be {\iid} Gaussian with covariance matrix $\tilde{\bm{\Upsilon}}$. The new measurement model incorporating the PMU measurements is given by
\begin{eqnarray}\label{eq:new_model}
\left[
\begin{array}{c}
\bm{y}\\
\tilde{\bm{y}}\\
\end{array}
\right] = \left[
\begin{array}{c}
\bm{f}(\bm{s})\\
\tilde{\bm{f}}(\bm{s})\\
\end{array}
\right] + \left[
\begin{array}{c}
\bm{v}\\
\tilde{\bm{v}}\\
\end{array}
\right]
\end{eqnarray}
The corresponding Jacobian matrix is given by
\begin{eqnarray}
\bm{J}_{\text{new}} = \left[
\begin{array}{c}
\frac{\partial(\bm{f}(\bm{s}))}{\partial(\bm{s})}\\
\frac{\partial(\tilde{\bm{f}}(\bm{s}))}{\partial(\bm{s})}\\
\end{array}
\right].
\label{eq:new_jacobian}
\end{eqnarray}
The error covariance matrix of the new estimate $\hat{\bm{s}}_{\text{new}}$ is $\bm{R}_{\hat{\bm{s}}_{\text{new}}} = \left(\bm{J}_{\text{new}}^{\mathrm{T}}\bm{\Upsilon}_{\text{new}}^{-1}\bm{J}_{\text{new}}\right)^{-1}$, where the noise covariance matrix
\[\bm{\Upsilon}_{\text{new}} = \left[ \begin{array}{cc}
\bm{\Upsilon} & 0 \\
0 & \tilde{\bm{\Upsilon}}\\
\end{array} \right].\]
The WLS estimate of $\bm{s}$ at the $(t+1)^{\text{th}}$ iteration, with PMU measurements, is $\hat{\bm{s}}_{t+1} = \hat{\bm{s}}_{t} + \bm{L}_{\text{new}, t}\bm{J}_t^{\mathrm{T}}\bm{\Upsilon}^{-1}\left(\bm{y} - \bm{f}(\bm{s}_{t})\right) + \bm{L}_{\text{new}, t}\tilde{\bm{J}}_{t}^{\mathrm{T}}\tilde{\bm{\Upsilon}}^{-1}\left(\tilde{\bm{y}} - \tilde{\bm{f}}(\bm{s}_{t})\right)$, where the gain matrix $\bm{L}_{\text{new}, t} \triangleq \left[\bm{J}_t^{\mathrm{T}}\bm{\Upsilon}^{-1}\bm{J}_t + \tilde{\bm{J}}_t^{\mathrm{T}}\tilde{\bm{\Upsilon}}^{-1}\tilde{\bm{J}}_t\right]^{-1}$, and the Jacobian matrix $\tilde{\bm{J}}$ is given by the partial derivative $\tilde{\bm{J}} \triangleq \frac{\partial(\tilde{\bm{f}}(\bm{s}))}{\partial(\bm{s})}$.

\vspace{-0.1in}
\subsection{Fault detection}\label{subsec:fault_detection}
As before, we model a power system with $N$ buses and $K$ branches modeled using an undirected graph $\mathcal G = \left(\mathcal V, \mathcal E \right)$, where $\mathcal V \triangleq \left\{1,\dots,N \right\}$ denotes the set of buses and $\mathcal E \triangleq \left\{(i_1, j_1),\dots,(i_K, j_K)\right\}$ denotes the set of transmission lines. We assume that no bus is connected to itself. As an example, we consider the IEEE-14 bus network with 14 buses and 21 branches. The voltage phasor angles at time $t$ are collected in the $N$-length vector $\bm x(t)$, $t = 1,\dots,T$. The active power flow between the buses at time $t$ is given by the $N \times N$ skew symmetric matrix $\bm Y(t)$, $t = 1,\dots,T$. We denote by $\bm B$ the $N \times N$ susceptance matrix on the transmission lines; $\bm{B}_{i,j} = \bm{B}_{j,i}$ is the susceptance between the buses $i$ and $j$ if $(i, j) \in \mathcal E$ and $\bm{B}_{i,j} = 0$ otherwise. We employ the linear DC power flow to describe the energy flow through each transmission line:
\begin{eqnarray}
\bm{Y}_{i,j}(t) = \bm{B}_{i,j}\left[\bm{x}_{i}(t) - \bm{x}_{j}(t) \right], t = 1,\dots,T.
\label{eq:dc_power_flow}
\end{eqnarray}

We let the $K$-length vector $\bm b$ with elements $\bm{B}_{i,j}$ for $(i, j) \in \mathcal E$ and $i > j$. Similarly, the elements $\bm{Y}_{i,j}(t)$ for $(i, j) \in \mathcal E$ and $i > j$ are collected in the $K$-length vector $\bm{z}(t)$. Thus, \eqref{eq:dc_power_flow} can be expressed as $\bm{z}(t) = \text{diag} \{\bm{b}\}\bm{D}\bm{x}(t)$, where $\text{diag} \{\bm{b}\}$ is the $K \times K$ diagonal matrix with the elements of the vector $\bm{b}$ along its main diagonal. The $K \times N$ matrix $\bm D$ is defined as follows: For the $k^{\text{th}}$ transmission line $(i_k, j_k) \in \mathcal E$ and $i_k > j_k$, we let $\bm{D}_{k, i_k} = 1$ and $\bm{D}_{k, j_k} = -1$, while the other elements in the $k^{\text{th}}$ row of $\bm{D}$ assigned zero. However, in practice, measurements of the power and the voltage phasor angles are noisy especially when one deals with real-time monitoring of the electric grid, {\ie}, we observe
\begin{eqnarray}
\tilde{\bm{z}}(t) = \bm{z}(t) + \bm{w}_z(t),\label{eq:noisy_powerflow} \\
\tilde{\bm{x}}(t) = \bm{x}(t) + \bm{w}_x(t), \label{eq:noisy_phasorangles}
\end{eqnarray}
where $\left\{\bm{w}_z(t)\right\}_{t=1}^{T}$ and $\left\{\bm{w}_x(t)\right\}_{t=1}^{T}$ are independent Gaussian noise processes with $\bm{w}_z(t) \sim \mathcal{N}\left(\bm{0}, \sigma_{z}^{2}\bm{I} \right)$, $\bm{w}_x(t) \sim \mathcal{N}\left(\bm{0}, \sigma_{x}^{2}\bm{I} \right)$, where $\bm{0}$ is the vector of all zeros, $\bm{I}$ denotes the identity matrix, and $\sigma_{z}^{2}$ and $\sigma_{x}^{2}$ are the noise variances.

Taking observations over $T$ time units, with a sampling rate of one sample/unit-time, the observation model can be compactly represented in matrix form as follows:
\begin{eqnarray}
\nonumber \bm{Z} &=& \text{diag}\{\bm{b}\}\bm{D}\bm{X}, \\
\tilde{\bm{Z}} &=& \bm{Z} + \bm{W}_{z},\label{eq:eiv_model}\\
\nonumber \tilde{\bm{X}} &=& \bm{X} + \bm{W}_{x},
\end{eqnarray}
where the dimensions of the matrices $\bm{Z}$ and $\bm{W}_z$ are $K \times T$, while those of matrices $\bm{X}$ and $\bm{W}_x$ are $N \times T$. This is commonly referred to as the linear errors-in-variables (EIV) model \cite{Huffel2002}. We consider the problem of detecting changes in the susceptance vector $\bm{b}$ based on the noisy observations $\tilde{\bm{Z}}$ and $\tilde{\bm{X}}$. Specifically, we assume knowledge of a vector $\bm{b}_0$ corresponding to the nominal behavior of the grid and pose a binary hypotheses testing problem:
\begin{eqnarray}
\begin{cases}
H_0: \bm{b} = \bm{b}_0, \\
H_1: \bm{b} \neq \bm{b}_0.
\end{cases}
\label{eq:test}
\end{eqnarray}
Under both hypotheses, $\bm{Z}$ and $\bm{X}$ are unknown and have to be estimated. A statistical model for \eqref{eq:eiv_model} was proposed in \cite{Wiesel2006} - \nocite{Wiesel2008}\cite{Beck2010} by letting $\bm{x}(t)$ be random variables centered around $\tilde{\bm{x}}(t)$. We obtain the maximum likelihood estimates of $\bm{Z}$ and $\bm{X}$ and use it in the context of detection to derive a hypothesis test termed TML-GLRT. The full details of TML-GLRT were first developed in \cite{Wei2012}. In the TML model, the observed $\tilde{\bm{x}}(t)$ are deterministic, while $\bm{x}(t)$ are random, {\ie}, $p(\bm{x}(t); \tilde{\bm{x}}(t)) \sim \mathcal{N}\left(\tilde{\bm{x}}(t), \sigma_x^2\bm{I} \right)$. Conditioned on $\bm{x}(t)$, we have $p(\tilde{\bm{x}}(t)|\bm{x}(t);\bm{b}) \sim \mathcal{N}\left(\text{diag}\{\bm{b}\}\bm{D}\bm{x}(t), \sigma_z^2\bm{I} \right)$. Therefore, the marginal distribution of $\tilde{\bm{z}}(t)$ is $p(\tilde{\bm{z}}(t);\bm{x}(t),\bm{b}) \sim \mathcal{N}\left(\text{diag}\{\bm{b}\}\bm{D}\tilde{\bm{x}}(t), \bm{H}(\bm b) \right)$, where $\bm{H}(\bm b) \triangleq \sigma_z^2\bm{I} + \sigma_x^2\text{diag}\{\bm{b}\}\bm{D}^{\mathrm{T}}\bm{D}
\text{diag}\{\bm{b}\}$. The TML-GLRT is given by
\begin{eqnarray}
\text{t}_{\text{TML}} = \log \frac{\max_{\bm{b}}\prod_{t=1}^{T}p\left(\tilde{\bm{z}}(t); \tilde{\bm{x}}(t); \bm{b}\right)}{\prod_{t=1}^{T}p\left(\tilde{\bm{z}}(t); \tilde{\bm{x}}(t), \bm{b}_0\right)} \stackrel[H_0]{H_1}{\gtrless} \gamma_{\text{tml}}, \label{eq:tml_glrt}
\end{eqnarray}
where $\gamma_{\text{tml}}$ is a fixed threshold. $\text{t}_{\text{TML}}$ can be written in the following simplified form:
\begin{eqnarray}
\nonumber \text{t}_{\text{TML}}\!\!\!\!\!\! &=& \!\!\!\!\!\! \max_{\bm{b}}\sum_{t=1}^{T}\log p\left(\tilde{\bm{z}}(t); \tilde{\bm{x}}(t); \bm{b}\right)
- \sum_{t=1}^{T}\log p\left(\tilde{\bm{z}}(t); \tilde{\bm{x}}(t), \bm{b}_0\right) \\ \nonumber
&=& \frac{1}{2}\text{Tr}\left\{\bm{A}^{\mathrm{T}}(\bm{b}_0)
\bm{H}^{-1}(\bm{b}_0)\bm{A}(\bm{b}_0) \right\} + + \frac{T}{2}\log |\bm{H}(\bm)| \\ && \!\!\!\!\!\! -
\min_{\bm b}\text{Tr}\left\{\bm{A}^{\mathrm{T}}(\bm b)\bm{H}^{-1}(\bm b)\bm{A}(\bm b) \right\} + \frac{T}{2}\log |\bm{H}(\bm)|,
\label{eq:tml_glrt_compact}
\end{eqnarray}
where $\bm{A}(\bm{b}) \triangleq \tilde{\bm{Z}} - \text{diag}\{\bm{b}\}\bm{D}\tilde{\bm{X}}$. The test \eqref{eq:tml_glrt_compact} compares the test statistic $\text{t}_{\text{tml}}$ to the threshold $\gamma_{\text{tml}}$, respectively. We choose the hypothesis $H_0$ if the test statistic is less than the threshold and $H_1$ if it is greater than the threshold. Given a large number of samples, the asymptotic performance of TML-GLRT under the hypothesis $H_0$ is given by $2\text{t}_{\text{tml}} \sim \chi_{K}^2$, where $\chi_{K}^2$ is the Chi-squared distribution with $K$ degrees of freedom \cite{Kay1998}. This result is independent of the specific value of the unknown $\bm X$. A reasonable approach to choosing the threshold for a given false alarm rate $\alpha$ is $\gamma_{\text{tml}, \alpha} = \frac{1}{2}F^{-1}_{\chi_{K}^2}(\alpha)$, where $F^{-1}_{\chi_{K}^2}(.)$ is the cumulative distribution function of the Chi-squared distribution with $K$ degrees of freedom.

\vspace{-0.1in}
\section{Experiments and discussion}\label{sec:simulation}\vspace{-0.05in}
For state estimation, we consider IEEE 14, 30 and 57 bus systems, and 50 time-sampled intervals with successful load flows under different loading conditions. The measurements of voltage magnitudes, power injections and power transfers are corrupted with additive white Gaussian noise having zero mean and standard deviations of 0.1\% and 2\% of actual values for voltages and powers, respectively. The generator outputs are adjusted to avoid overloading the swing bus.

The following three experimental scenarios are considered: (1) Normal load operation: the variation in loads are made to follow a linear trend of 10\%, 20\% and 30\%, with a random fluctuation of 3\% over the 50-sample time interval, without considering bad data in the noisy measurements accumulated at the control center; (2) Presence of bad data: at time instant $t$ = 25, one active and one reactive power transfer element with random error is introduced, without any errors at other time instants; and (3) Sudden load change: a portion of the load at one $\mathit{P}\mathit{Q}-$bus is disconnected at time instant $t$ = 30 to simulate a sudden load change scenario. We performed $M$ = 100 Monte Carlo simulations, and for each simulation a new set of noisy measurements is included.

Following the setup in \cite{Zhou2006}, PMU measurements are assumed to be obtained from devices with calibrations to correct for current and transformer turns ratio errors. Phasor magnitude errors are assumed to have a standard deviation of 3\% of the actual measurement, while phase angles are assumed to be in error only due to time synchronization errors. Phase angle error is assumed to have a standard deviation of 0.02 degree.

We consider the following performance indices commonly used in power systems \cite{Shih2002}: (i) The estimation error at every time step, averaged over the set of simulations is given by $\varepsilon_{t} = \frac{1}{M}\sum_{m=1}^{M}\left|\hat{\bm{s}}_{t,m} - \bm{s}_{t,m} \right|$, where $\hat{\bm{s}}_{t,m}$ is the estimated vector at the $m^{\text{th}}$ simulation. (ii) The performance index for the measurement vector at time step $t$:
\begin{eqnarray}\label{eq:error_perf2}
P_{\text{index}, t} = \frac{\sum_{m=1}^{N}\left|\hat{\bm{y}}_{t,m} - \bm{y}_{t,m}^{\text{noisefree}} \right|}{\sum_{m=1}^{M}\left|\bm{y}_{t,m} - \bm{y}_{t,m}^{\text{noisefree}} \right|},
\end{eqnarray}
where, at the $m^{\text{th}}$ simulation, $\hat{\bm{y}}_{t,m}$ is the estimated measurement vector, $\bm{y}_{t,m}^{\text{noisefree}}$ is the noise-free measurement vector, and $\bm{y}_{t,m}$ is the noisy measurement vector at the control center.

\begin{table*}[t]
\centering
\begin{tabular}{ | c | c | c | c | c |}\hline
    IEEE bus system & WLS & WLS with PMUs (topology) & WLS with PMUs ES ($\mathrm{scenario}~1$) & WLS with PMUs ES ($\mathrm{scenario}~2$) \\ \hline
    14 & 5.6511 & 5.6615 & 4.5142 & 3.6457\\ \hline
    30 & 3.1009 & 2.2388 & 1.9734 & 1.6535\\ \hline
    57 & 4.7611 & 3.5512 & 2.2149 & 1.0091\\ \hline
    \end{tabular}
    \caption{Estimation error $\varepsilon \times 10^{4}$ p.u. of WLS and WLS with PMU measurements, during normal operating condition. ES: electrical structure-based.}
    \label{tab:estimation_error}
\end{table*}
\begin{table*}[t]
\centering
\begin{tabular}{ | c | c | c | c | c |}\hline
    IEEE bus system & WLS & WLS with PMUs (topology) & WLS with PMUs ES ($\mathrm{scenario}~1$) & WLS with PMUs ES ($\mathrm{scenario}~1$) \\ \hline
    14 & 0.4402 & 0.3803 & 0.3391 & 0.3102\\ \hline
    30 & 0.4797 & 0.4198 & 0.3900 & 0.3596\\ \hline
    57 & 0.5002 & 0.4293 & 0.4000 & 0.3201\\ \hline
    \end{tabular}
    \caption{Time averaged performance index $P_{\text{index}}$ (p.u.) of WLS and WLS with PMU measurements, during normal operating condition. ES: electrical structure-based.}
    \label{tab:Jindex}
\end{table*}
Two different scenarios in the experimental setup are studied. In the first scenario (labeled $\mathrm{scenario}~1$), the same number of PMUs are employed for both the topology- and electrical structure-based approaches. As observed in \secref{sec:main_results}, the optimal location of these PMUs are different. For instance, let us consider the case of IEEE 14 bus system as an example. As shown in \tabref{tab:bus_pmu}, the topology-based PMU placement solution yields a minimum of 4 PMUs, while the electrical structure-based solution suggests a minimum of 7 PMUs. Therefore, for $\mathrm{scenario}~1$, phasor measurements from only 4 PMUs are incorporated into the WLS algorithm for both the topology- and electrical structure-based PMU placement schemes.
\begin{note}
In order to conform to the constraint of complete network observability on the optimization setup \eqref{eq:pmuplacement}, the number of PMUs installed on the grid is given by \tabref{tab:bus_pmu}, however, one only utilizes phasor measurements from 4 PMUs. The phasor measurements from the remaining 3 PMUs is neglected/discarded by the WLS algorithm for $\mathrm{scenario}~1$ though they remain installed on the grid.
\end{note}
For the second scenario (labeled $\mathrm{scenario}~2$), the number of PMUs follows from the second and third columns of \tabref{tab:bus_pmu} for the topology- and electrical structure-based approaches, respectively. And, WLS is implemented with phasor measurements from these PMUs.

In \tabref{tab:estimation_error}, the estimation error averaged over time for the 14, 30 and 57 bus networks under normal operating conditions are tabulated. As shown, the performance of WLS improves by incorporating PMU measurements for all the bus networks. Of particular importance is the improvement in performance of WLS aided by the electrical structure-based PMU placement ($\mathrm{scenario}~1$) compared to WLS incorporated with phasor data from topology-based placement scheme; compare values in columns 3 and 4. The improvement is due to better characterization of the connectedness between buses in the network as provided by the electrical structure-based PMU placement. The best performance of WLS is observed when it is aided by phasor measurements from all the PMUs obtained from electrical structure-based placement scheme (scenario 2); this is tabulated in column 5.
\vspace{-0.025in}
Next, in \tabref{tab:Jindex}, the performance index given by \eqref{eq:error_perf2} averaged over time for all the three bus systems, under normal operating conditions are tabulated. Here again, an improvement is seen in the performance of WLS by incorporating PMU measurements. $\mathrm{Scenario}~1$ yields a better performance than the topology-based scheme, however, $\mathrm{scenario}~2$ offers the best performance.

\begin{figure*}[t]
\centering
\begin{subfigure}{.33\textwidth}
  \centering
  \includegraphics[height=1.75in,width=2.5in]{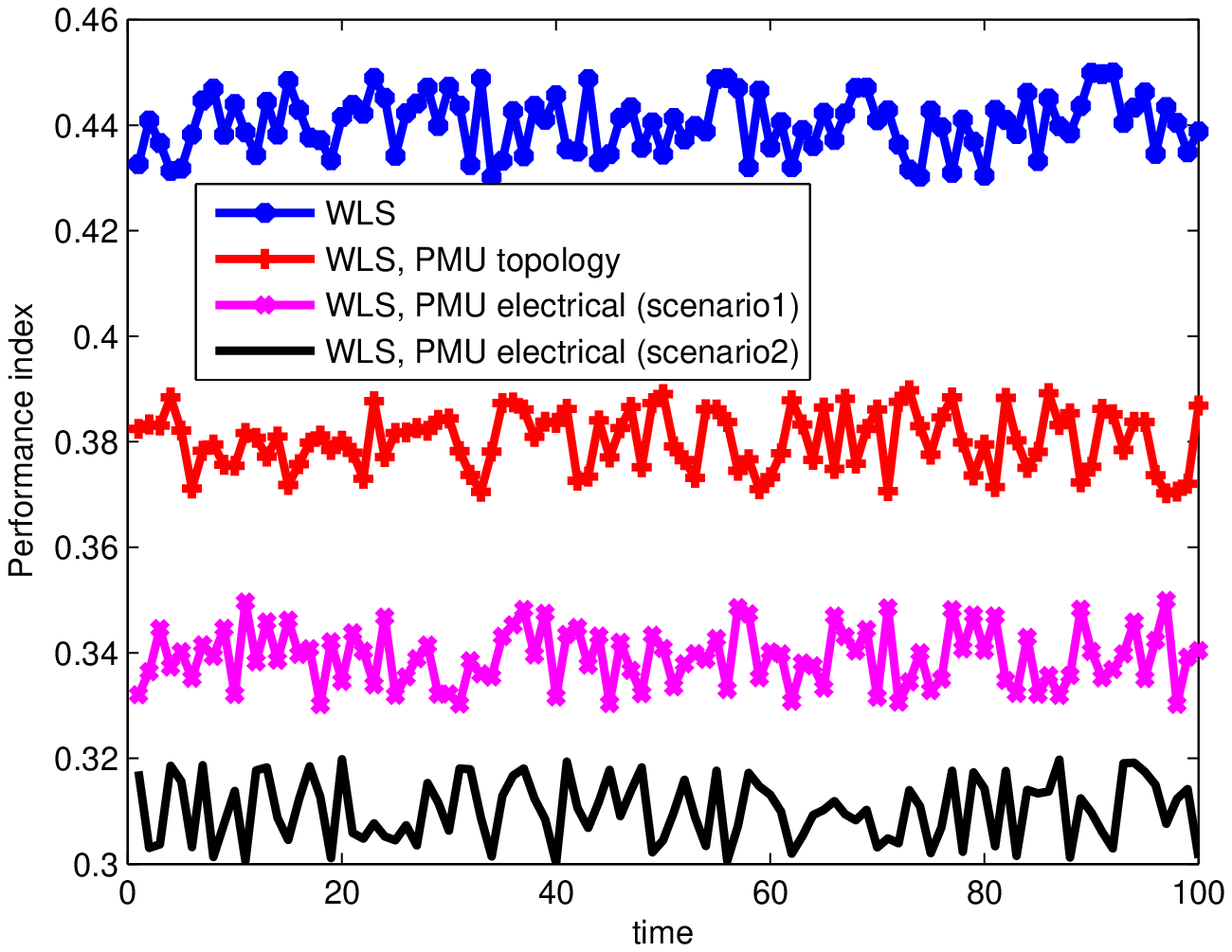}
  \caption{IEEE-14 bus system}
  \label{fig:normal14bus}
\end{subfigure}%
\begin{subfigure}{.33\textwidth}
  \centering
  \includegraphics[height=1.75in,width=2.5in]{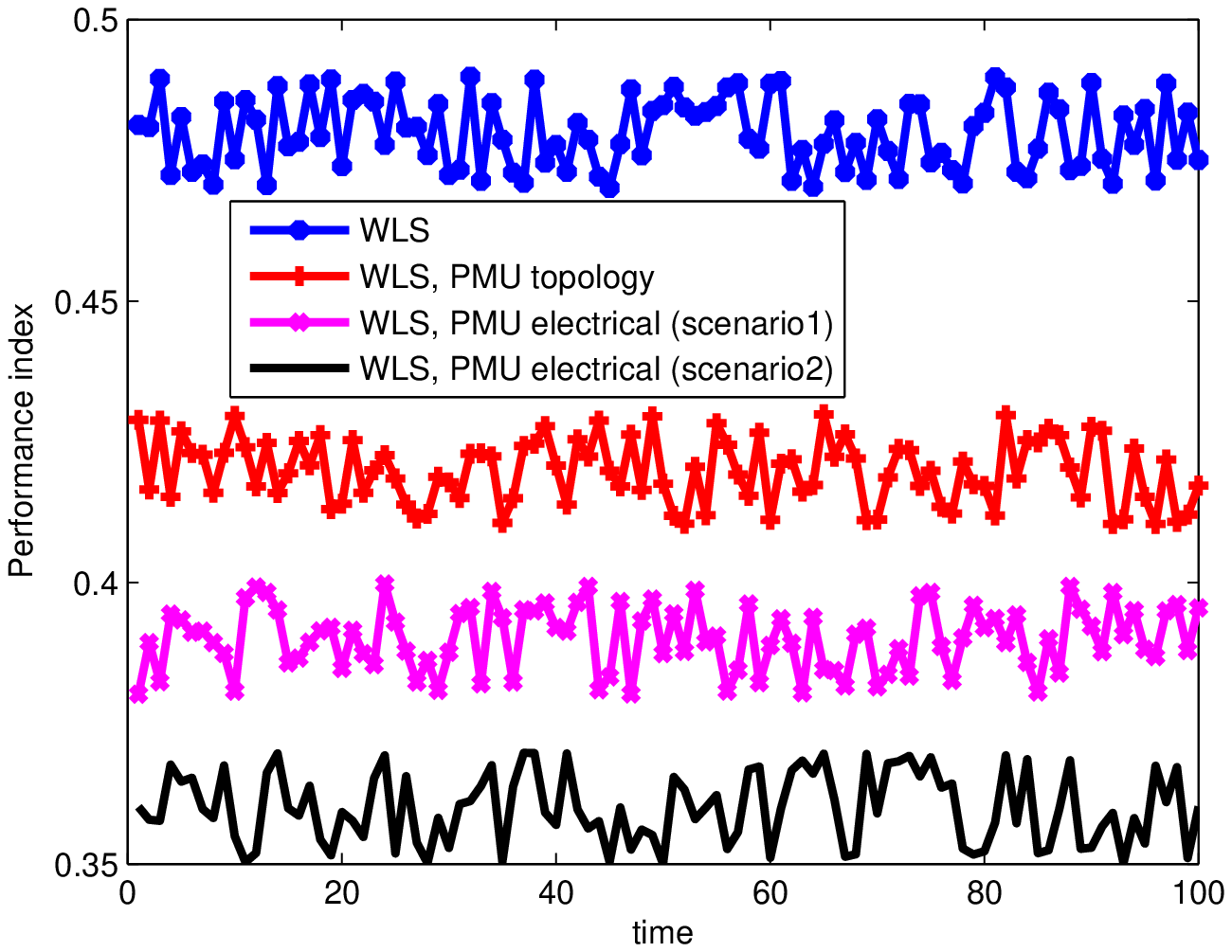}
  \caption{IEEE-30 bus system}
  \label{fig:normal30bus}
\end{subfigure}
\begin{subfigure}{.33\textwidth}
  \centering
  \includegraphics[height=1.75in,width=2.5in]{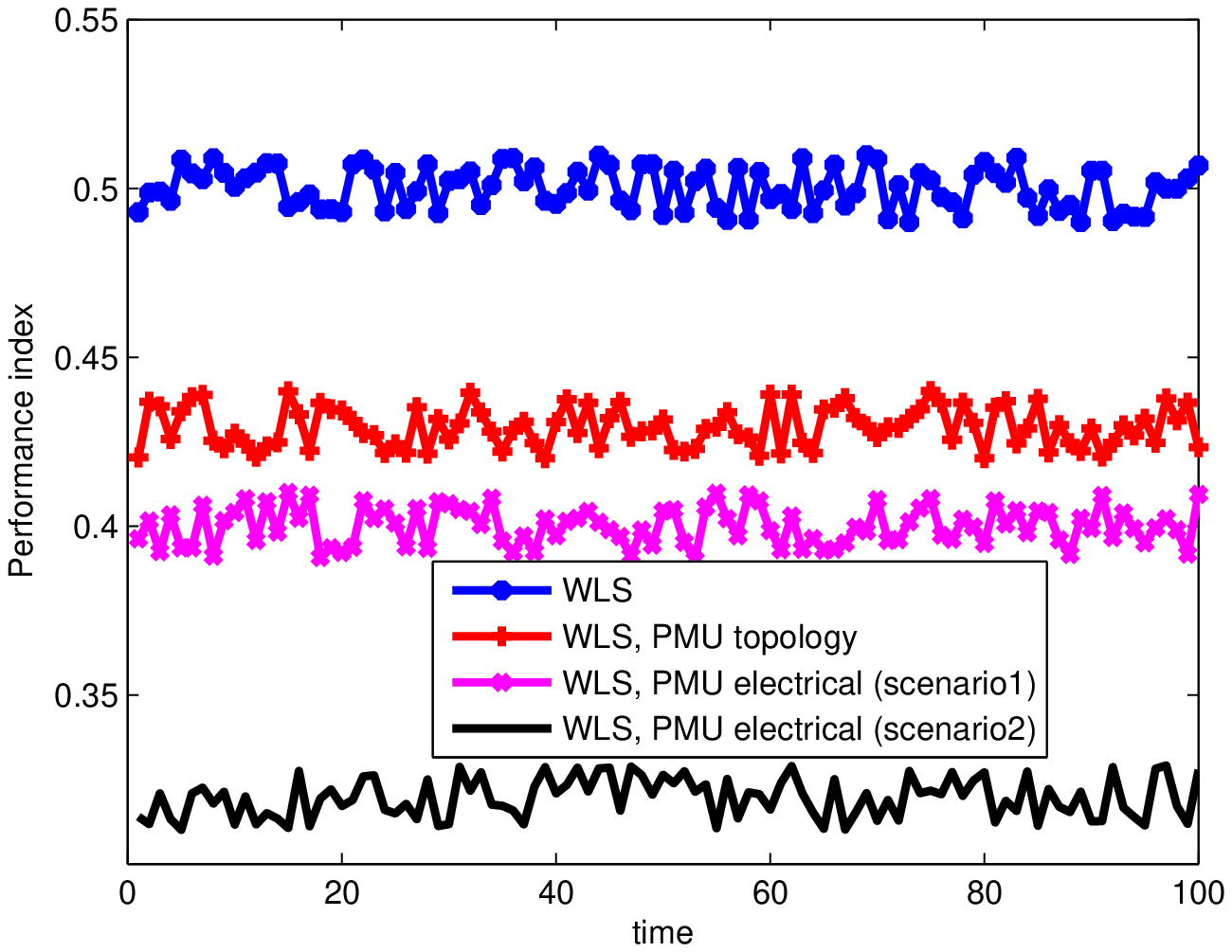}
  \caption{IEEE-57 bus system}
  \label{fig:normal57bus}
\end{subfigure}
\caption{Performance index under normal load conditions.}
\label{fig:performance_normal_load}
\end{figure*}
\begin{figure*}[t]
\centering
\begin{subfigure}{.33\textwidth}
  \centering
  \includegraphics[height=1.75in,width=2.5in]{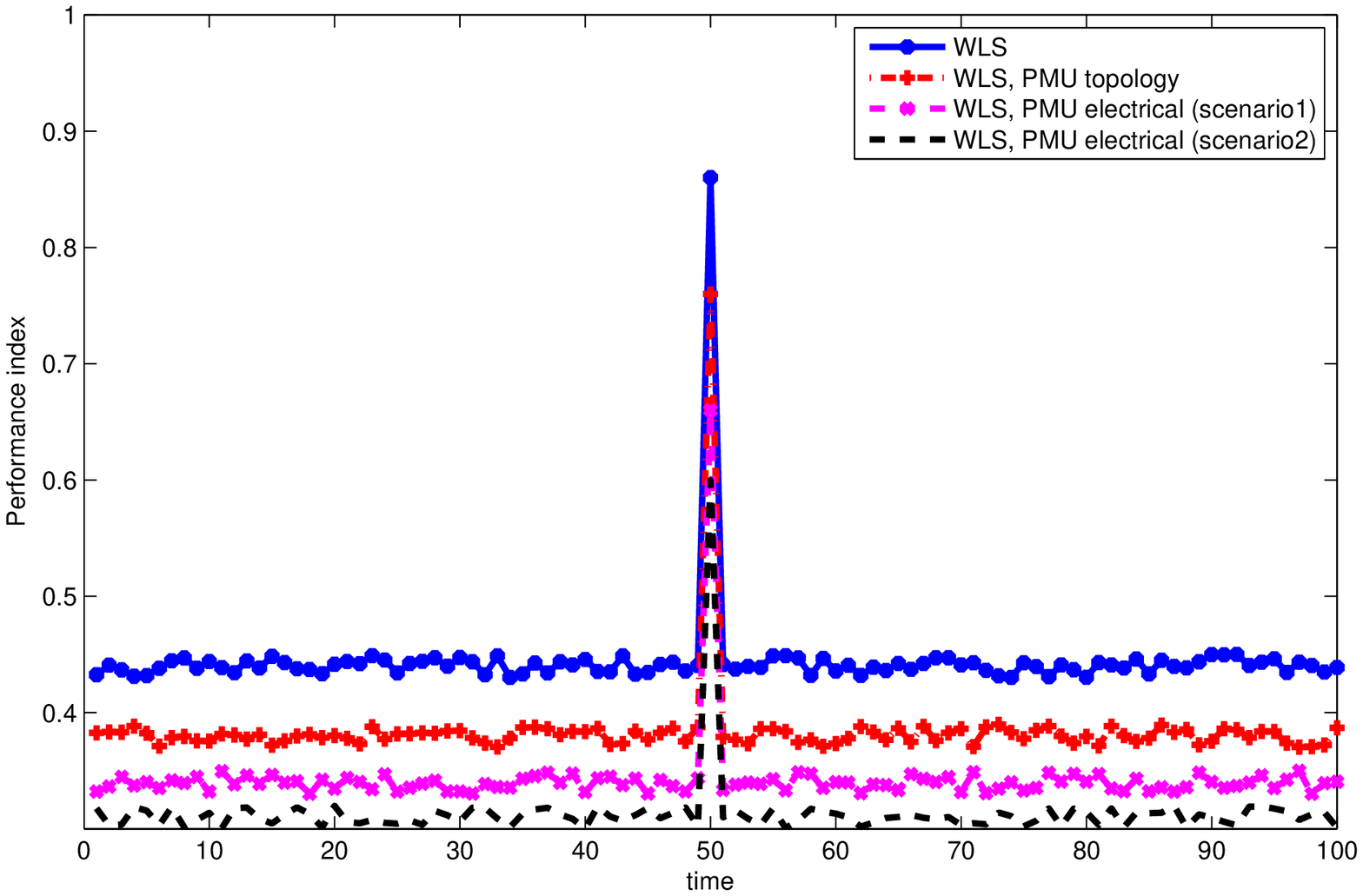}
  \caption{IEEE-14 bus system}
  \label{fig:baddata14bus}
\end{subfigure}%
\begin{subfigure}{.33\textwidth}
  \centering
  \includegraphics[height=1.75in,width=2.5in]{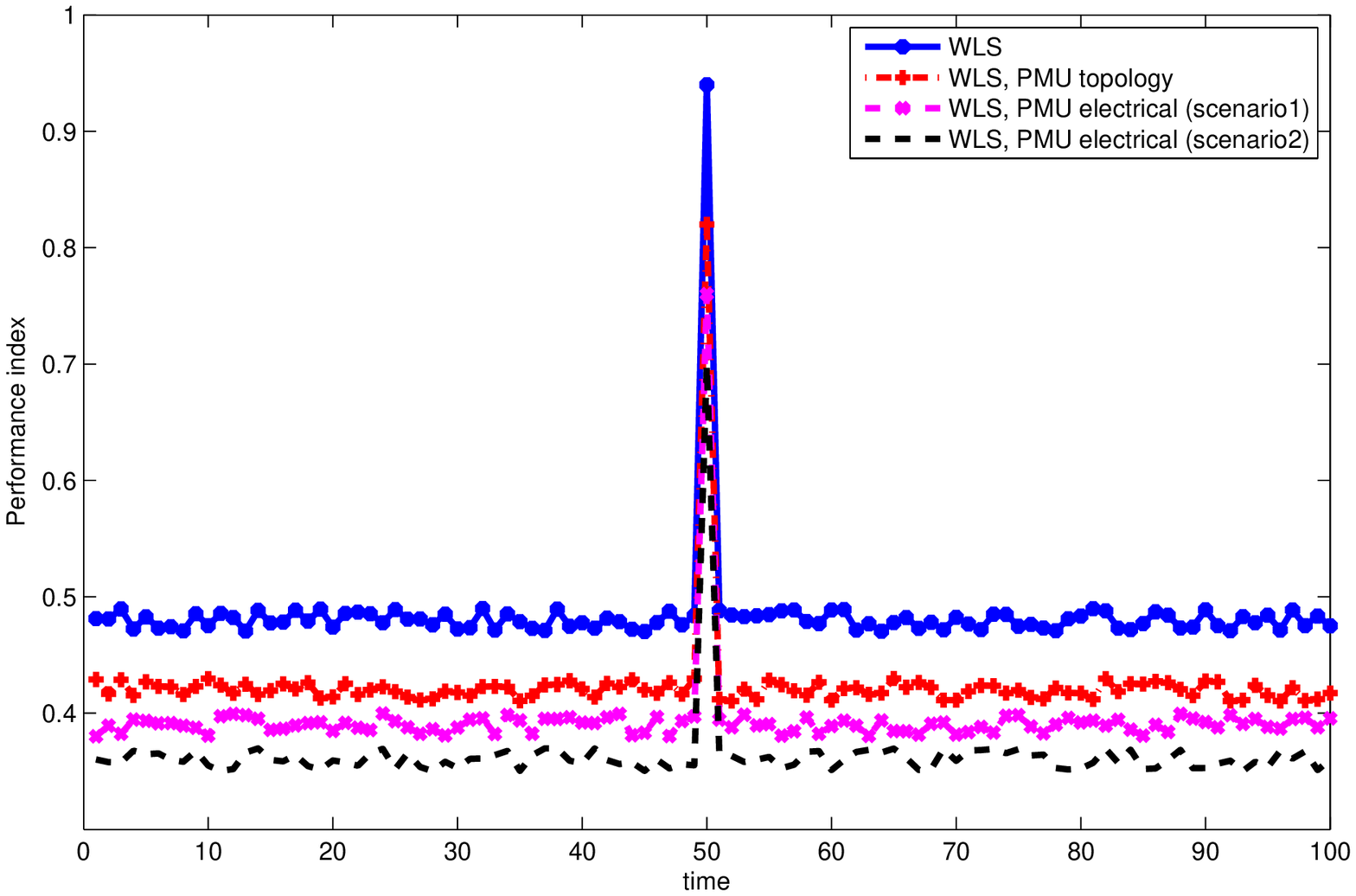}
  \caption{IEEE-30 bus system}
  \label{fig:baddata30bus}
\end{subfigure}
\begin{subfigure}{.33\textwidth}
  \centering
  \includegraphics[height=1.75in,width=2.5in]{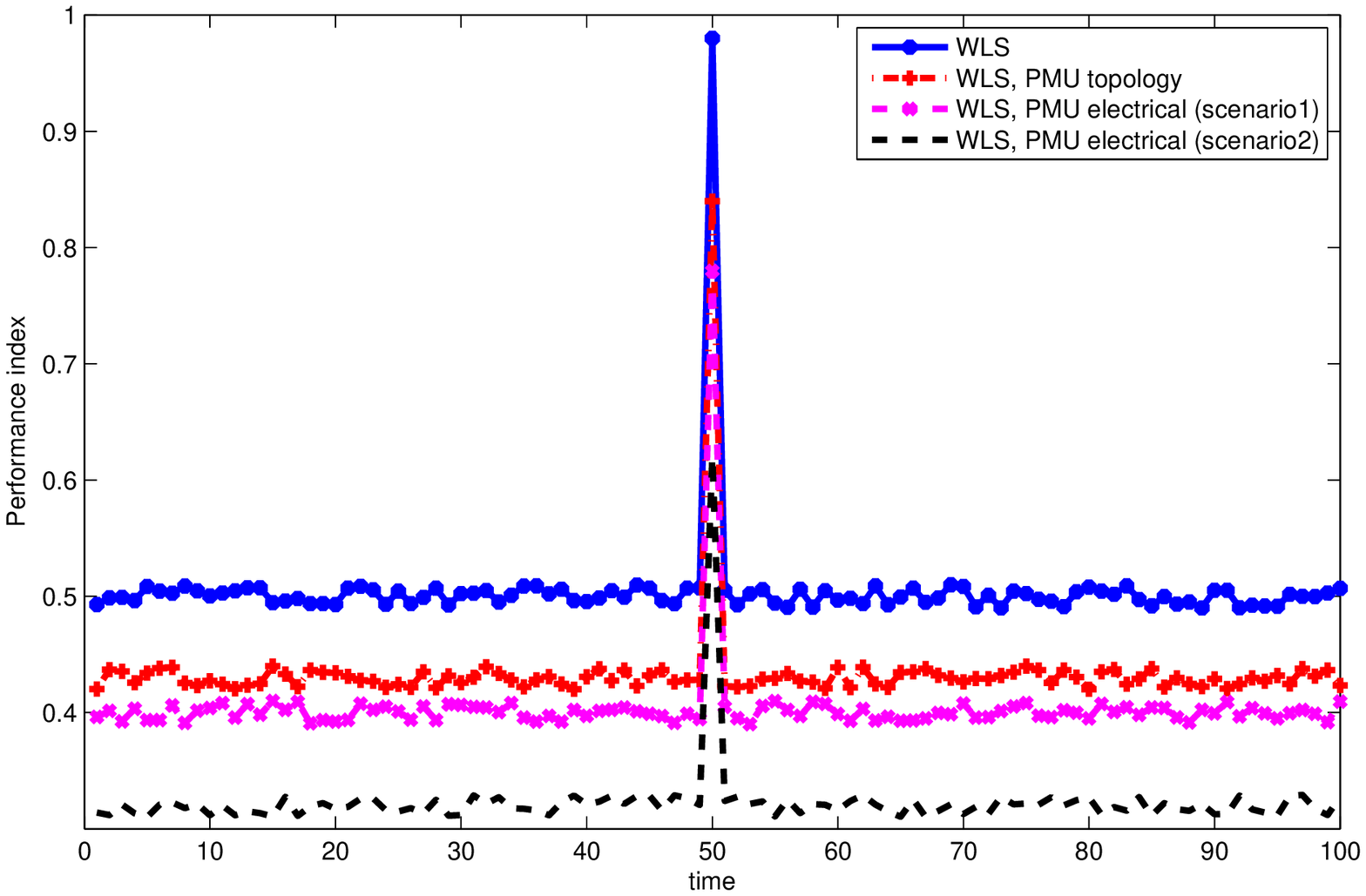}
  \caption{IEEE-57 bus system}
  \label{fig:baddata57bus}
\end{subfigure}
\caption{Performance index under bad data conditions.}
\label{fig:performance_bad_data}
\end{figure*}
\begin{figure*}
\centering
\begin{subfigure}{.33\textwidth}
  \centering
  \includegraphics[height=1.75in,width=2.5in]{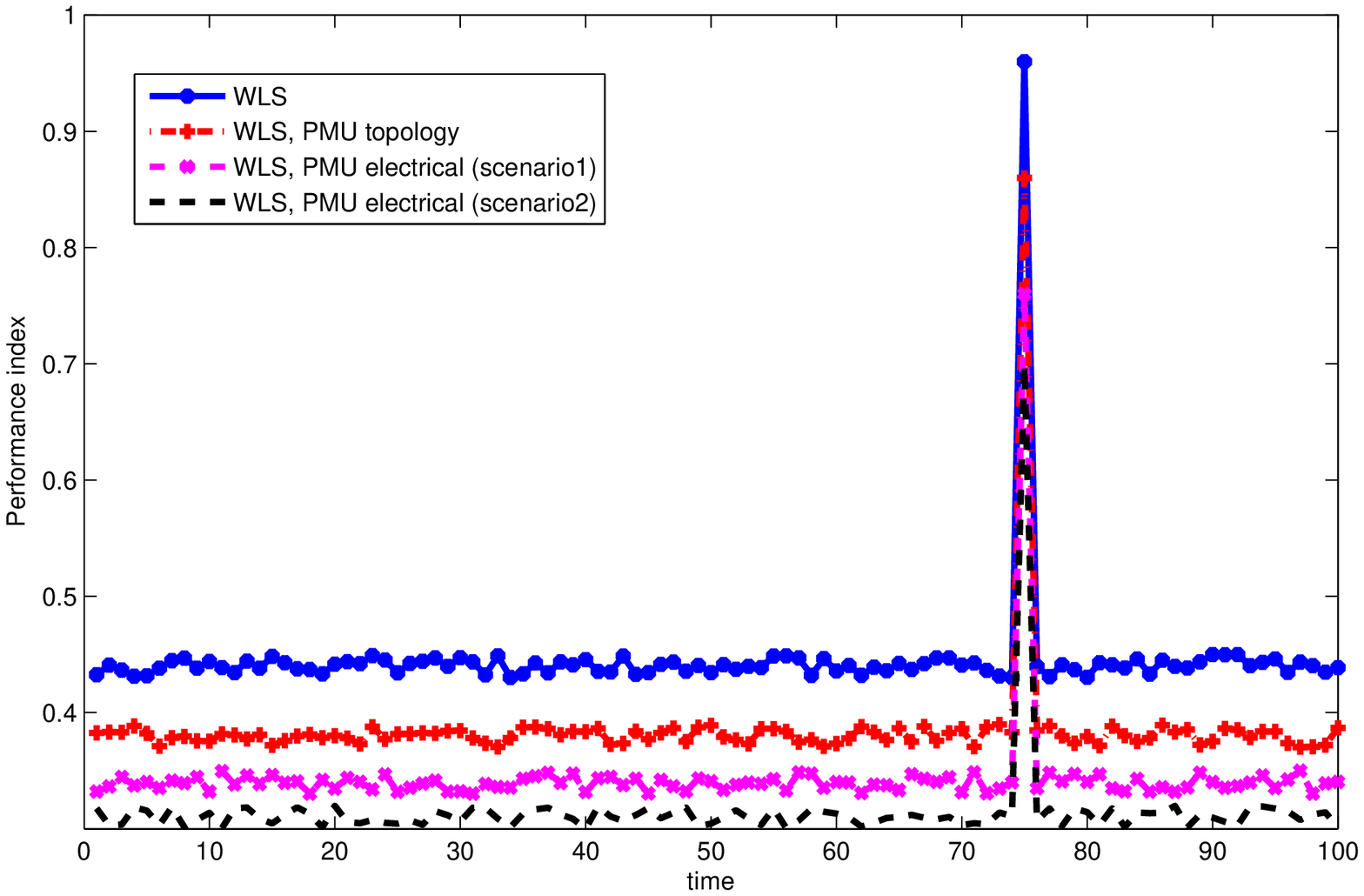}
  \caption{IEEE-14 bus system}
  \label{fig:suddenload14bus}
\end{subfigure}%
\begin{subfigure}{.33\textwidth}
  \centering
  \includegraphics[height=1.75in,width=2.5in]{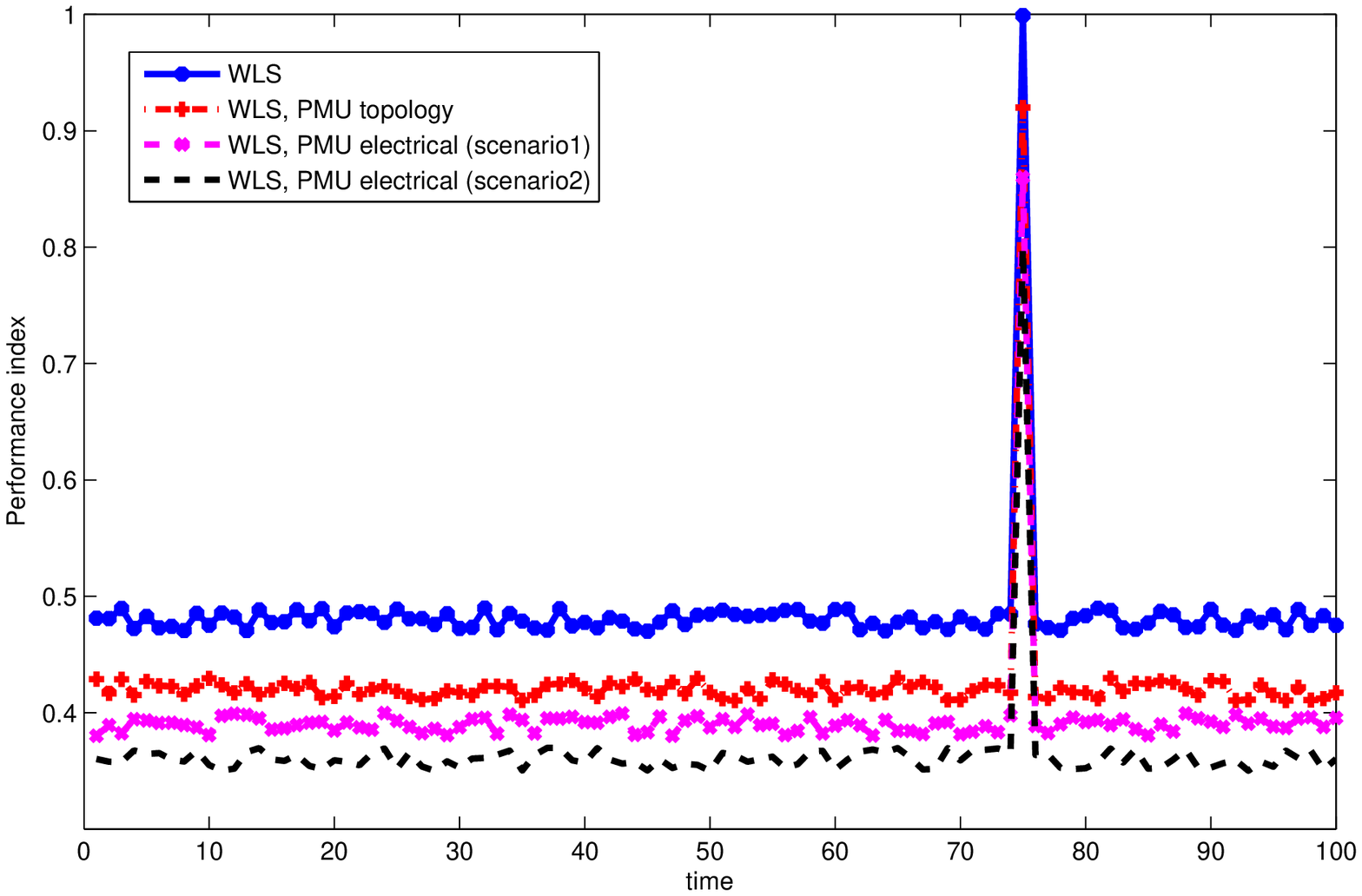}
  \caption{IEEE-30 bus system}
  \label{fig:suddenload30bus}
\end{subfigure}
\begin{subfigure}{.33\textwidth}
  \centering
  \includegraphics[height=1.75in,width=2.5in]{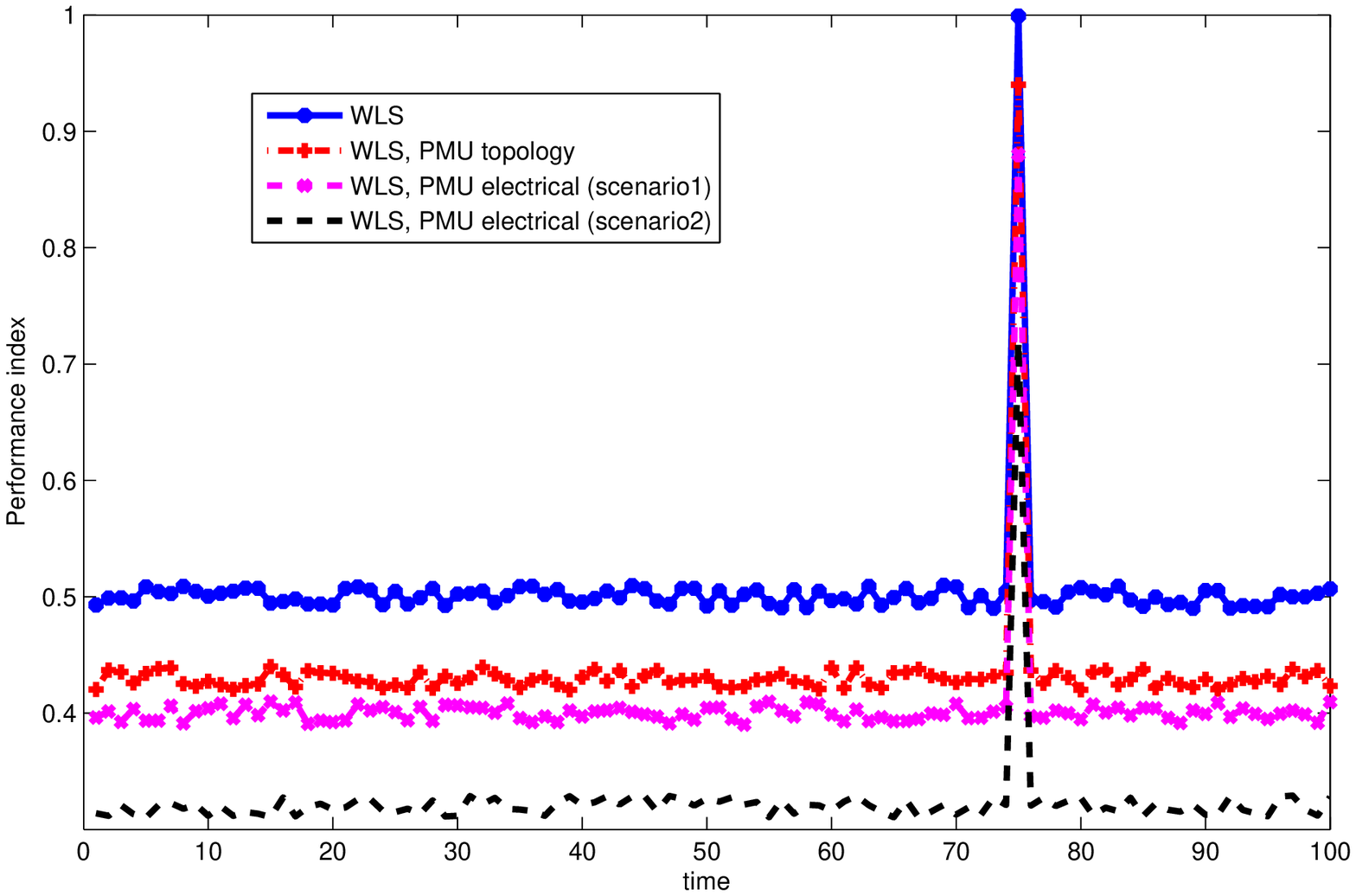}
  \caption{IEEE-57 bus system}
  \label{fig:suddenload57bus}
\end{subfigure}
\caption{Performance index under sudden load change.}
\label{fig:performance_sudden_load}
\end{figure*}
\vspace{-0.025in}
In \figref{fig:performance_normal_load}, the performance index $P_{\text{index},t}$ given by \eqref{eq:error_perf2} is plotted as a function of time for 100 time units while the system is operating under normal load conditions. In \figref{fig:performance_bad_data}, the plot of $P_{\text{index},t}$ is shown when the system has bad data injected at time instant $t$ = 50, while in \figref{fig:performance_sudden_load}, $P_{\text{index},t}$ is plotted when the system suffers from sudden load change at $t$ = 75. For all the three experiments, both $\mathrm{scenario}~1$ and $\mathrm{scenario}~2$ are considered. It is noticed that the performance of WLS shows an improvement when phasor data is incorporated into the algorithm. Also, in all the case studies, the electrical structure-based PMU placement scheme yields a superior performance compared to the topology-based approach.

To analyze the performance of TML-GLRT with PMU measurements, we consider the IEEE 14 test bus system. The noise variances are $\sigma_{z}^2$ = $\sigma_{x}^2$ = 0.01. We perform 10$^4$ Monte Carlo simulations. At each realization, the elements of $\bm{X}$ are generated independently as standard normal random variables. The active power flow $\bm{Y}$ between the buses were obtained using the power flow algorithm in MATPOWER \cite{Zimmerman2011}; the susceptance vector $\bm{b}$ was then computed using $\bm{Y}$. Under $H_0$, we use the susceptance parameters given by the test profile. Under $H_1$, we apply a change of -2\% to every element of $\bm{b}$. Here, -2\% essentially means that every element of $\bm{b}$ is made smaller by a factor of 2\%; this number can be arbitrarily chosen without loss of generality. We compute the variation of probability of detection $(P_d)$ for a change in the susceptance parameter versus time, using a fixed value of false alarm rate ($\alpha$) uniformly picked between [0, 0.2] for which the threshold $\gamma_{\text{tml}, \alpha}$ is calculated. This procedure is repeated for different values of $\alpha \in$ [0, 0.2] resulting in a set of $P_d$s versus time. We plot the average (over the number of $\alpha$) $P_d$ versus time.

\begin{figure*}
\centering
\begin{subfigure}{.33\textwidth}
  \centering
  \includegraphics[height=1.75in,width=2.5in]{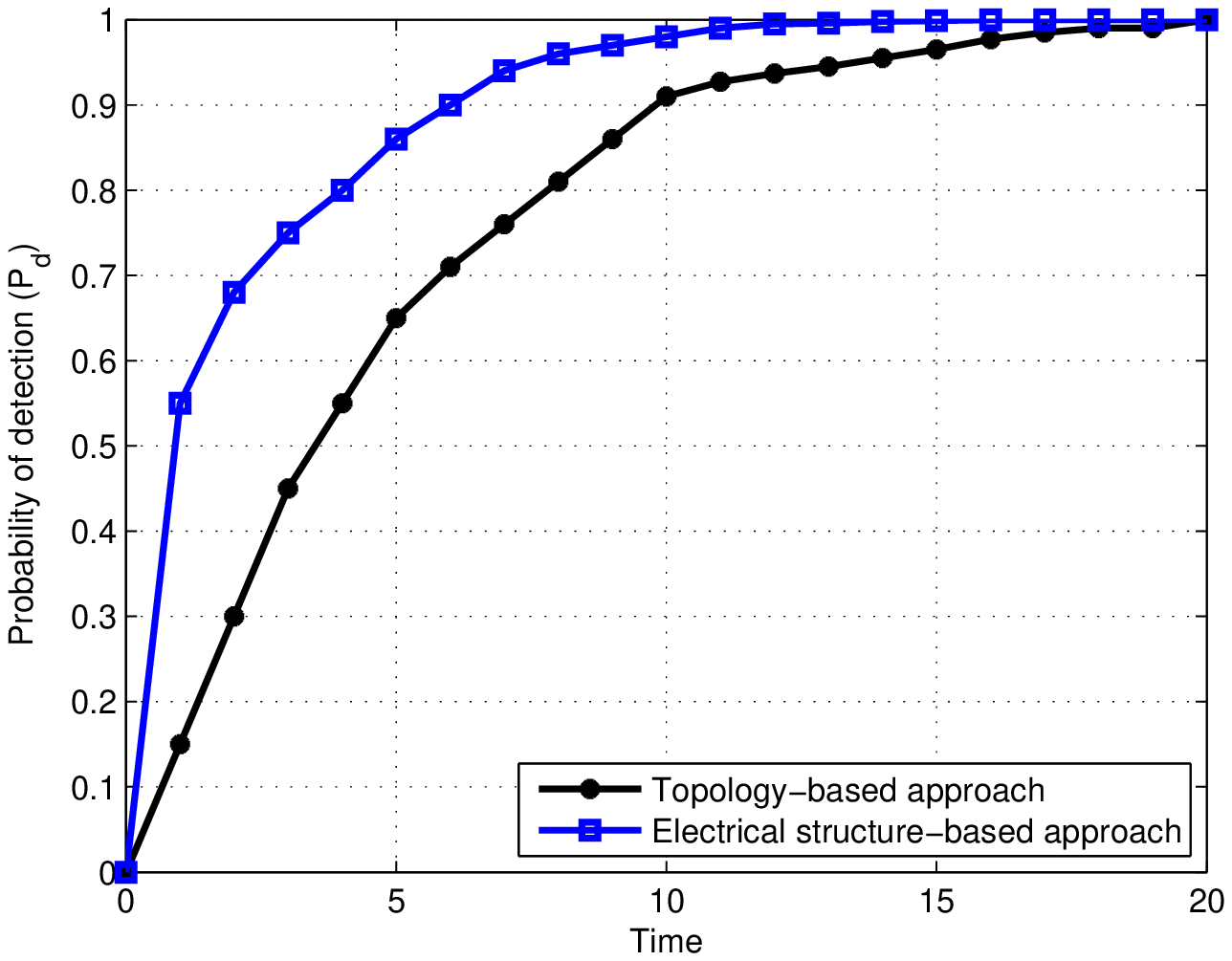}
  \caption{}
  \label{fig:pd_time_topo_elec1}
\end{subfigure}%
\begin{subfigure}{.33\textwidth}
  \centering
  \includegraphics[height=1.75in,width=2.5in]{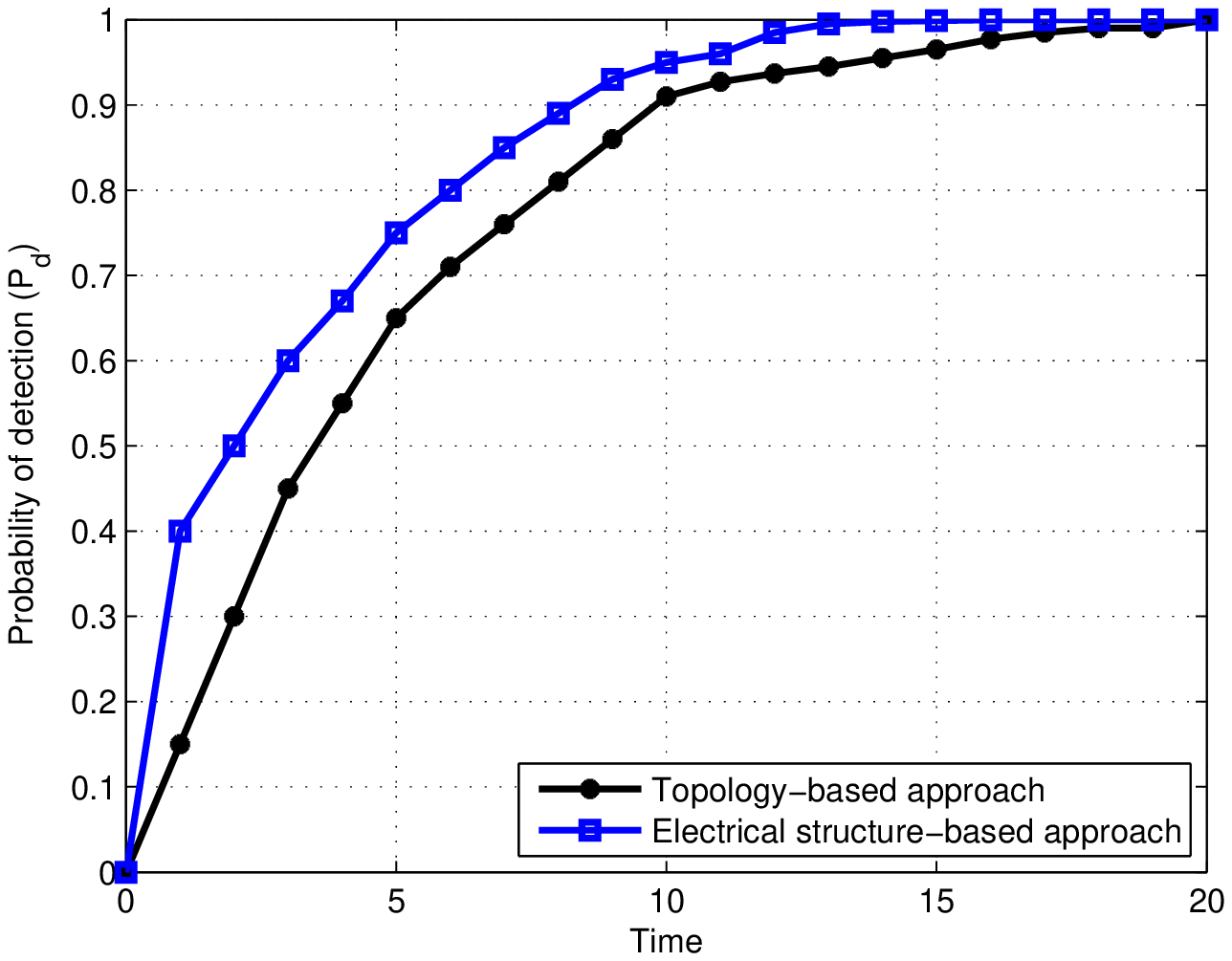}
  \caption{}
  \label{fig:pd_time_topo_elec2}
\end{subfigure}
\begin{subfigure}{.33\textwidth}
  \centering
  \includegraphics[height=1.75in,width=2.5in]{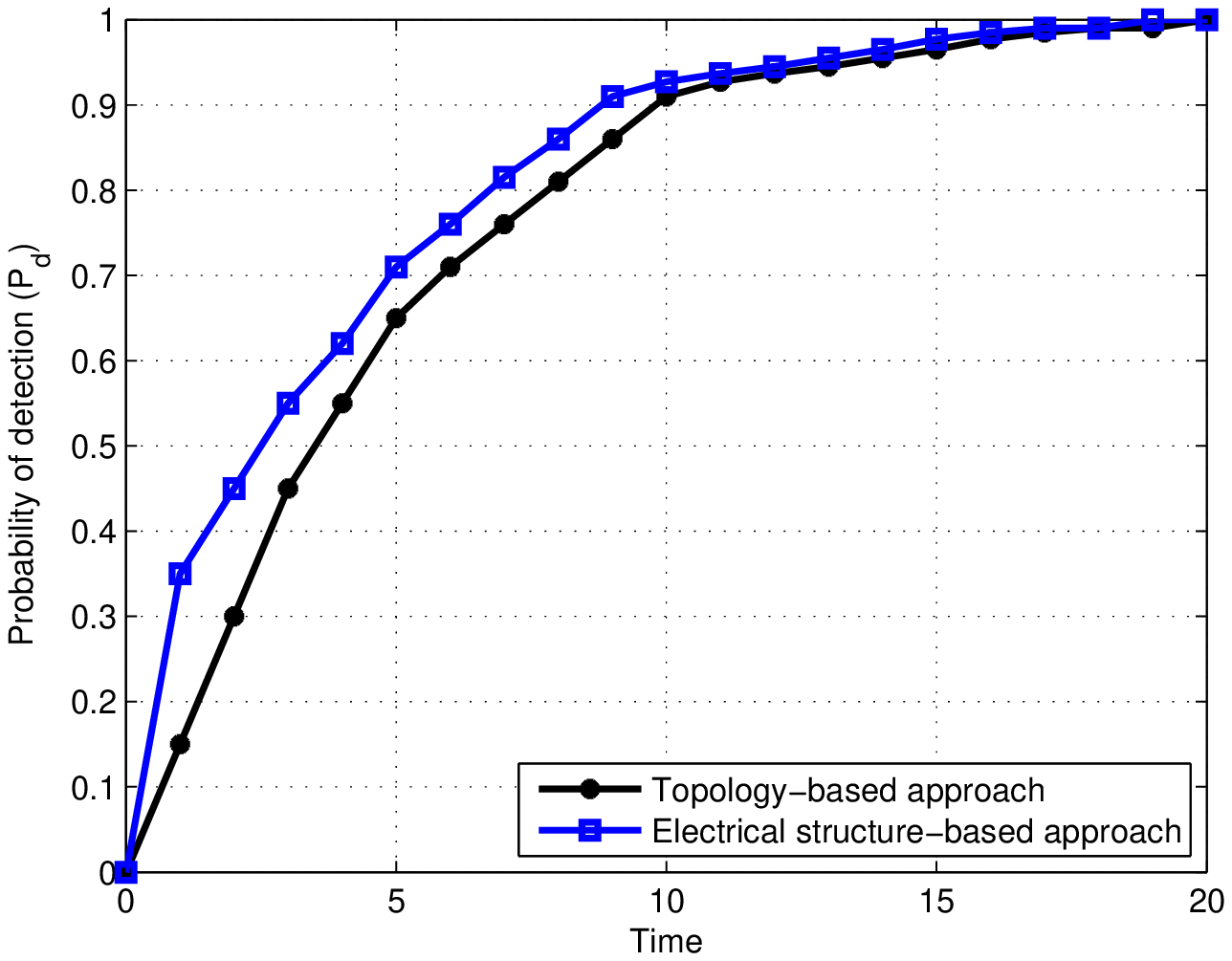}
  \caption{}
  \label{fig:pd_time_topo_elec3}
\end{subfigure}
\caption{Probability of detection versus time, for one frame lasting 20 time units, for IEEE-14 bus system.}
\label{fig:pd_time}
\end{figure*}
For the topology-based approach, we obtain an optimum 4 PMUs to be installed on buses 2, 6, 7 and 9. For the electrical structure-based approach, we need an optimum number of 7 PMUs to be installed on buses 1, 3, 8, 11, 12, 13 and 14. The plots of $P_d$ versus time for one frame lasting 20 time units when the PMUs are placed employing both the approaches are shown in \figref{fig:pd_time_topo_elec1}. For a fairer comparison, we modify the experimental setup as follows: We employ the same number of PMUs for both the topology and electrical structure-based approaches to analyze the detection performance. Specifically, we employ only 4 PMUs transmit their phasor data to the control center for both PMU placement approaches for fault detection. More precisely, for the topology-based approach we allow PMUs on buses numbered 2, 6, 7 and 9 to transmit as before. For the electrical structure-based approach, we now allow only 4 PMUs on buses numbered 8, 11, 12 and 14 transmit their phasor data. This choice was based on the electrical centrality measure between various buses on which the PMUs are placed; this was characterized in an earlier work by the authors \cite{Nagananda2014a}. As seen in \figref{fig:pd_time_topo_elec2}, the electrical structure-based approach has a higher detection performance compared to it topological counterpart.

Finally, we conduct an experiment in which we allow lesser number of PMUs for the electrical structure-based approach to transmit compared to the topology-based approach. Specifically, we allow only 3 PMUS (on buses 14, 8 and 12, based on the based on the electrical centrality measure) to transmit for the electrical structure-based approach, while continuing with 4 PMUs for the topology-based approach. The resulting plots of $P_d$ versus time are shown in \figref{fig:pd_time_topo_elec3}, where it is clearly seen that $P_d$ for the electrical structure-based approach is slightly higher than the topology-based approach.

\begin{note}\label{note:fig5_6}
For \figref{fig:pd_time_topo_elec2} and \figref{fig:pd_time_topo_elec3}, it should be noted that we have installed the optimum number of 7 PMUs for the electrical structure-based approach, so as to achieve complete network observability. However, we allowed only 4 PMUs to transmit their phasor data to the control center. Though the remaining 3 PMUs record phasor data across the lines on which they are installed, they do not transmit them to the control unit. So long as phasor data from the ``high priority'' ({\ie}, higher electrical centrality measure) buses are obtained, the test statistic can be computed for hypothesis testing.
\end{note}

The electrical structure-based approach provides a stronger characterization of the electrical influence between network components compared to the topology-based approach. Therefore, the probability of detection of change in the susceptance parameter is higher when the PMUs are installed based on the electrical structure-based approach compared to the case when they are installed using the topology-based approach.
\vspace{-0.1in}
\section{Conclusion}\label{sec:conclusion}\vspace{-0.07in}
The PMU placement problem was addressed from the perspective of the electrical structure of the grid. PMU measurements were incorporated into WLS used for static state estimation and TML-GLRT used to detect changes in the susceptance parameters of the grid. The main result was that on the one hand, a larger number of PMUs were required to achieve complete network observability, while on the other hand, it improved the performance of practical applications such as state estimation and fault detection. Given that the electrical structure provides a more comprehensive description of interconnections in the power network, our results promote a realistic perspective of PMU placement in the grid.

\vspace{-0.07in}
\bibliographystyle{IEEEtran}
\bibliography{IEEEabrv,proposals}
\raggedbottom

\end{document}